\documentclass[10pt,conference]{IEEEtran}
\usepackage{subcaption}
\usepackage{multirow}
\usepackage{graphicx}
\usepackage{url}
\usepackage{bbm}
\usepackage{xcolor}
\usepackage{makecell}

\begin{document}

\title{Ten Ways in which Virtual Reality Differs from Video Streaming}
\author{\IEEEauthorblockN{Gustavo de Veciana$^{\dagger}$, Sonia Fahmy$^{\ddagger}$, George Kesidis$^{*}$, Voicu Popescu$^{\ddagger}$}
\IEEEauthorblockA{$^{\dagger}$\textit{The University of Texas at Austin} \;
$^{\ddagger}$\textit{Purdue University} \; $^{*}$\textit{Penn State University}}}

\maketitle

\begin{abstract}
Virtual Reality (VR) applications have a number of unique characteristics that set them apart from traditional video streaming. These characteristics have major implications on the design of VR rendering, adaptation, prefetching, caching, and transport mechanisms. This paper contrasts VR to video streaming, stored 2D video streaming in particular, and discusses how to rethink system and network support for VR.
\end{abstract}

\section{Introduction}
\label{sec:intro}

Virtual Reality (VR) is a powerful form of interactive visualization that provides \textit{immersion}, by covering the entire field of view (FoV) of the user, \textit{depth perception}, by catering distinct images, with appropriate disparity, to the user's left and right eyes, and a \textit{natural interface} for specifying the desired view, by tracking the user's head. VR users can \textit{interact} with elements of the \textit{virtual environment} (e.g., a vehicle or an instrument) or with \textit{each other} in unpredictable ways. 
\textcolor{black}{VR is expected to serve as the foundation for the ``metaverse''~\cite{metaverse}. For example, the company Meta web site states that ``Like the internet, the metaverse will help you connect with people when you aren't physically in the same place and get us even closer to that feeling of being together in person.''}

Consider the scenario of a civil engineering course whose students visualize and interact with a large-scale urban simulation in VR. The urban virtual environment with thousands of buildings, with vegetation, and with vehicle and pedestrian traffic may be too large to download to, or to render on, a headset, e.g., Meta's Quest~3. While most of the students may be in a lecture room, some may attend the course remotely. The students may be near each other in the virtual environment, for example when following the instructor as a group, or may be in disparate locations, when allowed to explore the virtual environment individually. Whereas some of the virtual environment data does not change, such as that modeling the buildings, some data changes in real time, such as the position and orientation of the student avatars and of objects with which the students interact. The students should be able to easily switch to a different simulation or to a different application altogether as they progress through the lecture or from one course to the next. The headsets alone cannot support such a VR application satisfactorily. They require support from a server that may be remote, i.e., in the \emph{cloud}, and/or nearby, i.e., at the \emph{edge}, and a network that connects the headset client device to the server(s). This network must have a Wi-Fi or cellular last hop for an untethered experience (to avoid tripping hazards while physically walking, for example). 

Since headsets are now at usability, performance, and price points suitable for mass adoption,
VR has the potential to revolutionize healthcare, education, training, and entertainment. However, realizing its potential for widespread societal impact is hindered by the lack of effective network and system VR support. Although there have been significant advances in system and network support for traditional 2D video streaming~\cite{netflix-talk}, these advances do not easily translate to VR. This is because VR has unique user requirements and uses large and heterogeneous virtual environment models that render traditional support methods unsuitable. \textcolor{black}{There is a wide gap between the computer graphics research community where new VR data representations are being developed, and the computer networking community, much of which adapts techniques developed for 2D video streaming~\cite{flare,vivo,GROOT20,M5,MuV2}.}

In this paper, we highlight ways in which networked virtual reality differs from video streaming. Although we focus on pre-recorded (stored) 2D video streaming, we note that VR also differs from stored panoramic (360\textdegree) video streaming, stored volumetric (free viewpoint) video streaming, \textcolor{black}{live video streaming}, and video conferencing. Panoramic video streaming~\cite{flare,CoREpaper} is, in many ways, simpler than VR since user interaction is limited to rotational movement of the user's FoV (i.e., 3 degrees of freedom). Although volumetric video streaming~\cite{vivo,GROOT20,M5,MuV2} allows both rotational and translational movement (i.e., 6 degrees of freedom), neither panoramic nor volumetric video streaming allows users to interact with each other or with the virtual environment
synchronously in real time. These interactions can be complex, resource and data intensive, and difficult to predict. \textcolor{black}{Live video streaming significantly differs from VR-- one difference is again the limited interaction.} 
Finally, VR differs from video conferencing in that, despite both being real time, some elements of the virtual environment may be static and can therefore be prefetched or cached (in addition to again the limited interaction). In summary, the unique characteristics of VR require fundamentally rethinking the design of its system and network support.

Table~\ref{tbl:comparison} summarizes ten key differences between stored 2D video streaming and networked VR. We group the ten differences into four major categories: (A) application characteristics, (B) rendering and adaptation, (C) prefetching and caching, and (D) transport. The remainder of this paper discusses each of these differences. 

\begin{table*}[htb]
\centering
\caption{Key differences between stored 2D video streaming and networked VR.}
\begin{tabular}{|l|p{2in}|p{3.5in}|} 
\hline
\textbf{Aspect} & \textbf{Stored 2D Video Streaming} & \textbf{Networked Virtual Reality}\\ 
\hline\hline
\multicolumn{3}{|c|}{\textbf{(A) Application characteristics}}\\
\hline
1. Interaction 
& \parbox[m]{2in}{Mostly sequential viewing, with typically-infrequent viewing-position changes.}
& \parbox[m]{3.5in}{\vspace*{2pt} Users change location and orientation, and potentially interact with the virtual environment or other users in real time and in unpredictable ways; users may teleport. \vspace*{2pt}}  
\\
\hline
2. Content dynamics
& \parbox[m]{2in}{Mostly MPEG-encoded video (with potentially additional streams for encoded audio and subtitles).}
& \parbox[m]{3.5in}{\vspace*{2pt}
Heterogeneous and may include (textured) polygon meshes or point clouds, animations, stored or live video, audio, and subtitles. Much of the content may be static and may have a long useful life. \vspace*{2pt}}
\\
\hline
3. Resource needs 
& \parbox[m]{2in}{Relatively simple due to predictable asynchronous interaction with each user.}
& \parbox[m]{3.5in}{\vspace*{2pt} Complex due to an individualized experience, heterogeneity of content, and synchronous interaction. When a set of users is close (in both the virtual environment and the physical world), can amortize computation, caching, and communication costs. \vspace*{2pt}}\\ 
\hline
4. QoE requirement 
& \parbox[m]{2in}{QoE is a function of startup delay, stalls, frame quality and variability.}
& \parbox[m]{3.5in}{\vspace*{2pt} Requires $\sim$10~ms motion-to-photon latency to prevent cybersickness. QoE is a complex function of usability, 3D perception, response time to every action, delay with every teleportation, headset energy consumption and temperature, in addition to frame quality and variability.\vspace*{2pt}} 
\\
\hline\hline
\multicolumn{3}{|c|}{\textbf{(B) Rendering and adaptation}}\\
\hline
5. Rendering 
& \parbox[m]{2in}{Video displayed on client; no computation on an edge server is typically required.}
& \parbox[m]{3.5in}{\vspace*{2pt} Choice between (i) Rendering on client device (constrained by device capacity and does not leverage multi-client sharing) or (ii) rendering on an edge server (constrained by network latency and bandwidth), or (iii) partial rendering on both client and server(s).\vspace*{2pt}}
\\
\hline
6. Adaptation 
& \parbox[m]{2in}{Select one of a few available bitrates for each (typically 4-second) ``chunk.''}
& \parbox[m]{3.5in}{\vspace*{2pt} Dynamically employ level-of-detail and visibility-based virtual environment complexity reduction based on the user field of view.\vspace*{2pt}}
\\
\hline\hline
\multicolumn{3}{|c|}{\textbf{(C) Prefetching and caching}}\\
\hline
7. Prefetching 
& Sequentially prefetch as long as playout buffer space is available.
& \parbox[m]{3.5in}{\vspace*{2pt} Difficult to accurately predict user view and behavior for prefetching.\vspace*{2pt}} 
\\
\hline
8. Caching 
& \parbox[m]{2in}{Already-viewed content typically discarded \textcolor{black}{at the client}.}
&  \parbox[m]{3.5in}{\vspace*{2pt} Can cache parts of virtual environment that may be visited or revisited; cache replacement and sharing policies important.\vspace*{2pt}}
\\
\hline\hline
\multicolumn{3}{|c|}{\textbf{(D) Transport}}\\
\hline
9. Prioritization & \parbox[m]{2in}{Data is mostly fetched sequentially.}
& \parbox[m]{3.5in}{\vspace*{2pt} Prioritize data to fetch based on visibility, level of detail, and sharing.\vspace*{2pt}}
\\
\hline
10. Delivery 
& \parbox[m]{2in}{\vspace*{2pt} In-order, reliable, delivery to the application typical.\vspace*{2pt}}
& \parbox[m]{3.5in}{Do not always need in-order delivery or complete reliability.}
\\
\hline
\end{tabular}
\label{tbl:comparison}
\end{table*}

\section{Application Characteristics}

We begin by contrasting the interaction, content, and resource needs for video streaming and VR applications, and then discuss their Quality of Experience (QoE) requirements.

\subsection{Interaction, Content Dynamics, and Resource Needs}

A user often streams a stored video sequentially from start to finish, but there may be, typically infrequent, position changes (e.g., rewind, fast forward). 
Video is encoded and compressed using standards such as MPEG, VP9 and VC1. Separate streams may be used for audio, encoded using standards such as AAC, and closed captioning or subtitles~\cite{netflix-talk}.
Stored video streaming applications typically handle each user separately since users are not synchronized, i.e., may not be viewing the same time point of the video at the same time. Due to the asynchronous and predictable request/response interaction with each user, video streaming servers are stateless and can scale to large numbers of simultaneous users.

User interaction within immersive VR applications is much more complex and unpredictable. A VR user frequently changes location and orientation, and may interact with the virtual environment or with other users synchronously in real time. Further, VR users may occasionally teleport to different locations. 

\begin{figure}[htb]
\centering
        \includegraphics[width=0.9\columnwidth]{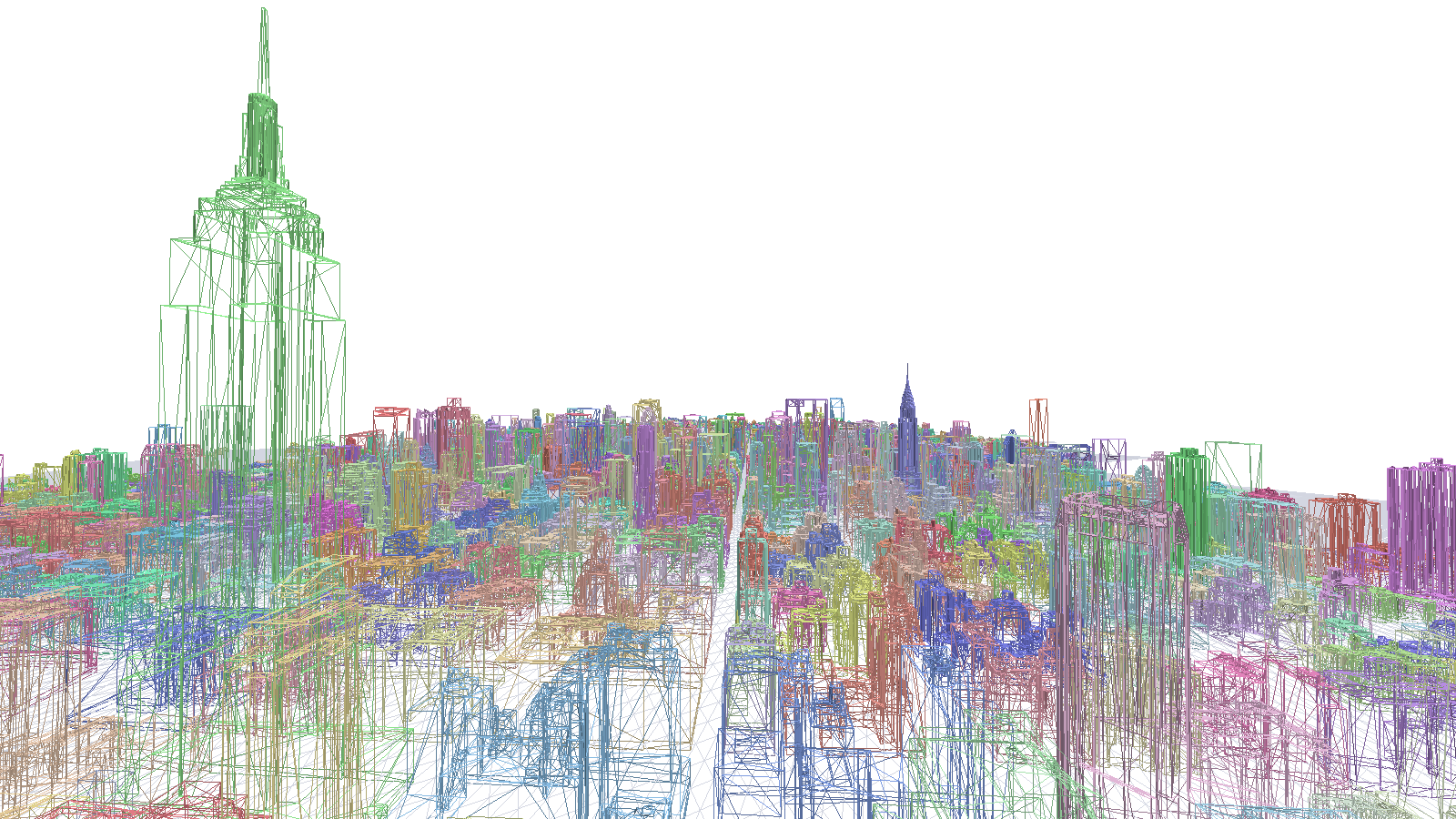}
        \includegraphics[width=0.9\columnwidth]{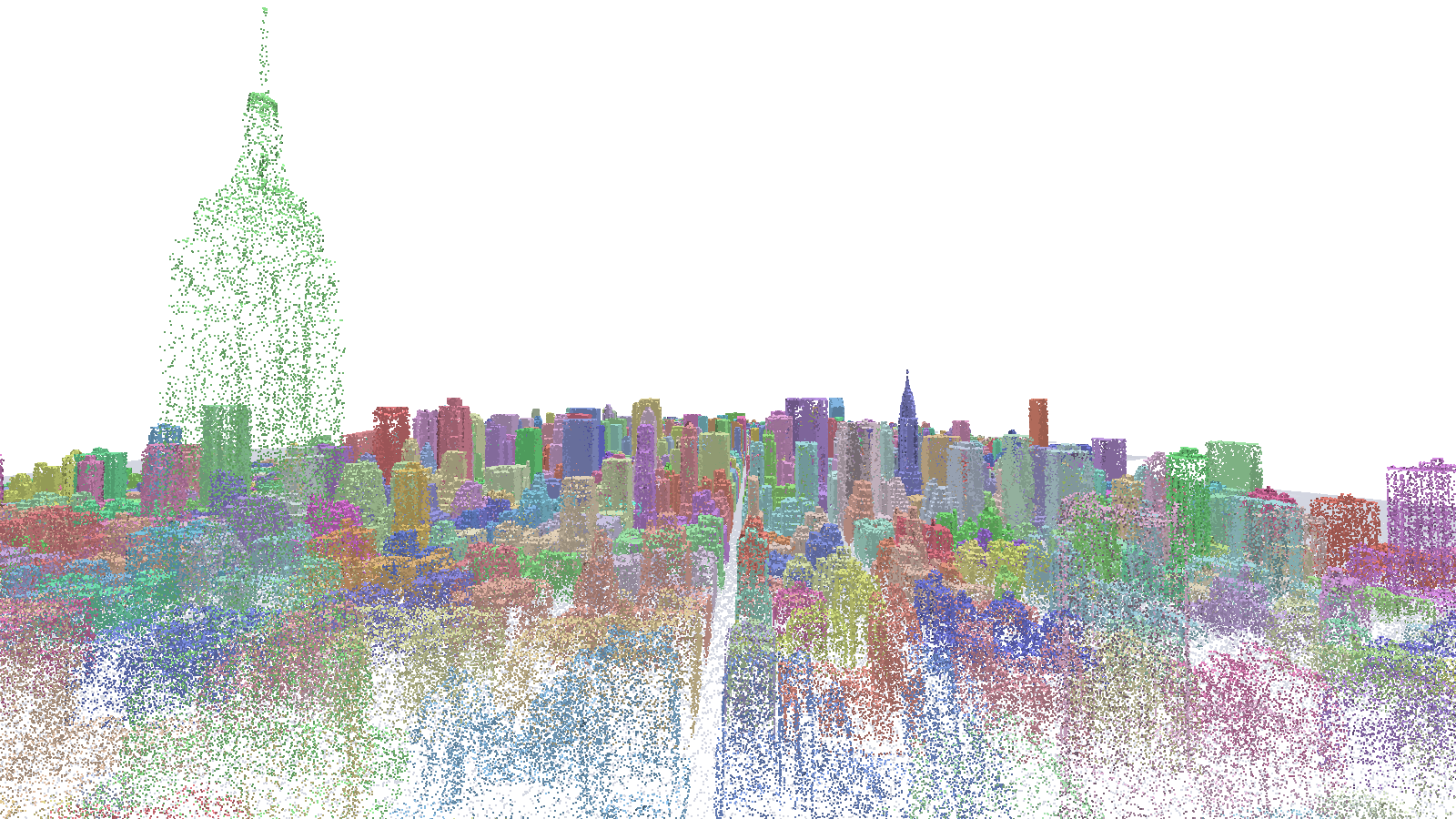}
        \caption{Manhattan urban virtual environment. The triangle mesh representation (top), shown here in wireframe, has 3.7 million triangles and 2 million vertices. The point-based representation (bottom), shown here with 2 $\times$ 2 pixel splats, has 20 million points and a density of 0.1 points/m$^2$.}
        \label{fig:meshAndpoints}
\end{figure}

Virtual environments are represented by heterogeneous data types, with different sizes, arrival deadlines, \textcolor{black} {complexity reduction strategies, rendering methods,} and useful life ranges. VR data typically includes static (textured) polygon meshes or point clouds, but may also include animations, stored or live video, audio, and subtitles. For example, Figure~\ref{fig:meshAndpoints} depicts a rendered polygon (triangle) mesh and a rendered point cloud of the same virtual environment. \textcolor{black} {A mesh encoding is more compact because the triangle vertices are the minimal set of 3D points needed to approximate the geometry. In contrast, the point cloud additionally includes points inside triangles to enforce a desired minimum point density. The less compact representation, implemented by the point cloud, offers simpler complexity management, because it enables the use of hierarchical space partitioning schemes that leverage the uniform distance between points, and it bypasses the challenge of having to modify the explicit connectivity data stored by meshes. However, the lack of explicit connectivity complicates rendering, in that the output frame footprint of individual points, i.e., the ``splat'' size, must be approximated based on the neighbors of a point.}

Meshes and point clouds may be static and have a long useful life, since the user may visit and revisit them multiple times while navigating through the virtual environment. However, a virtual environment may simultaneously include more dynamic types of data such as real-time audio, animated objects, or live video being projected on a stationary object in the virtual environment.

VR resource needs are complex due to the need to create an individualized and customized user experience. Creating such an experience requires processing heterogeneous data types and synchronization with other users. The user experience must be computed on the fly as it is not possible to pre-render the environment for all possible user trajectories, views, and interactions. This increases the dimensionality of the problem and makes scaling to a large number of users difficult. However, when a subset of VR users is in close proximity in both the virtual environment and the physical world (e.g., users on the same Wi-Fi network), it is desirable for the VR support platform to amortize computation, caching, and communication costs, as discussed throughout the remainder of this paper. 

\subsection{QoE Requirement}

In stored video streaming, user engagement can be inferred by simple metrics such as noting when the user stops watching the video or the frequency of restarts or pauses.
In the case of VR, \emph{usability} of a virtual environment can be more complex and is crucial to understanding user QoE. For example, if a user is unable to accomplish a task such as finding or selecting an object, or unable to see updates made by other users in a timely manner, this results in a degraded QoE~\cite{kim2020systematic}. 

For stored video streaming, it is important to minimize the startup latency and the duration of video stalls during rebuffering periods. In the case of VR, a short response time to user view changes and to user actions is crucial, as latency above 10~ms is a major factor leading to  cybersickness~\cite{Kourtesis19}. 
Quickly observing the effects of actions of other users is also important in multi-user settings. Further, instead of the one-time cost of startup latency in the case of video streaming, post-teleportation latency in VR can occur several times during a single session.

Objectively assessing the quality of a displayed frame is important for both video streaming and VR, and is a mature research area, e.g., using metrics such as Structural Similarity Index Measure (SSIM)~\cite{SSIM}. Relatively less attention has been paid to smoothness of quality changes~\cite{Bovik11}. For example, avoiding large quality variation spatially (over a frame) and temporally (across frames) may have a higher impact on user QoE than the overall quality of an individual frame. Transmitting and rendering an unnecessarily high level of detail can result in aliasing effects, not to mention waste scarce resources.

Resource consumption and temperature of a headset are also important factors in VR. Energy consumption impacts how long a user can use the application before needing to recharge the headset, and the temperature of the headset can impact user comfort while using the application. 

Determining the relative importance, or weights, to assign to each of the above factors to obtain a single number for user QoE remains a challenge. Some work has attempted to base such weight assignment on results of user studies~\cite{YuZu}, but it remains an open research problem.

Overall, the QoE of VR applications must reflect a much more dynamic environment than video streaming, wherein users may act in complex and unpredictable ways and frequently teleport from one location to another. The presence of such real-time transients driven by user viewing and interactive behavior, and the need to capture usability, make the problem of designing objective QoE metrics for VR challenging.

\section{Rendering and Adaptation}

Given the application characteristics discussed above, we now turn to their implications on the design of system and network support. We begin by exploring where and how rendering and adaptation to resource constraints take place.

In traditional video streaming, a video is temporally divided into ``chunks'' and transmitted to the client device. A chunk is typically four seconds in duration, but values of one or two seconds are also used. The video chunks are displayed on the client device without needing any computation on the server.

Video streaming applications adapt to changing network conditions using the Dynamic Adaptive Streaming over HTTP (DASH) protocol. DASH simply requests from the server one out of the few (typically no more than ten) available bitrates of each chunk of the video.  A standard HTTP web server can simultaneously serve a large number of client devices because it is completely stateless. These HTTP servers typically form a Content Distribution/Delivery Network (CDN) such as the Netflix Open Connect Network~\cite{netflix-talk}.

Unlike the case of video streaming, VR creates \emph{user-specific} visual content based on the user's current FoV and state on the fly. The process of creating (i.e., rendering) this user-specific visual content may be executed on the client, on the server, or on a combination of both. 

A pure \emph{client-based} rendering approach is limited by the size and complexity of virtual environment models that the client headset can handle (e.g., 1.5 million triangles for the Quest~3). Pure client-based rendering also consumes limited headset energy, and cannot take advantage of any potential sharing in multi-user settings. 

A pure \emph{server-based} remote rendering approach relieves clients from rendering duties by sending pre-rendered frames to the clients. However, this approach is typically impractical, as a per-frame request/response interaction with a server, even a nearby server, incurs latency that is unacceptable in the VR context. Further, the communication overhead may be excessive. 

Due to the limitations of pure client-based and pure server-based approaches, an approach that partitions rendering operations between a client and a server is advantageous. In this case, creating a VR experience involves communicating \emph{compact virtual environment data representations} to the client device, and rendering client-specific visualization (and audio) on the client device. Client devices may further leverage neural enhancement~\cite{YuZu} to increase the rendered quality.

Reducing the complexity of the virtual environment to create compact data representations and to adapt to limited network and system resources is a challenging problem that has been, and continues to be, the subject of significant research. The two most popular approaches for complexity reduction are Level of Detail (LoD) adaptation~\cite{luebke2003level} and visibility-based adaptation. LoD adaptation fetches a lower level of detail for (parts of) the virtual environment (typically distant from the user), with possible loss in QoE, whereas visibility-based adaptation performs view frustum culling and, in certain cases, occlusion-based culling~\cite{vivo}. These adaptation approaches are dependent on the user's current FoV, not just on resource constraints as in the case of 2D video streaming.

LoD adaptation and visibility-based adaptation algorithms differ according to the type of data representation used, e.g., polygon meshes, textured polygon meshes, point clouds, or panoramic videos. We give three examples in the remainder of this section.

\textbf{Example 1:} Consider LoD adaptation of a triangle mesh. LoD adaptation in this case requires finding a triangle mesh with fewer vertices and fewer triangles than the original mesh, yet, when seen from a distance, the coarser triangle mesh looks similar to the original mesh. LoD adaptation to meet a given triangle budget and to achieve a smooth transition between consecutive LoDs is a challenging geometric problem.

Some VR applications rely on an extreme form of LoD adaptation with only two LoDs: a high LoD for the part of the virtual environment close to the user, i.e., the \emph{near} region, and a low LoD for the part of the virtual environment far from the user, i.e., the \emph{far} region, e.g.,~\cite{coterie}. The high LoD corresponds to the original triangle meshes. The low LoD is implemented by pre-rendering the far region geometry to an \textit{environment map}, i.e., a 360$^\circ$ panorama encoded as a ``cubemap'' that serves as a backdrop. To minimize the wait time at startup and after teleportation, the near region starts out small and progressively grows. 

One advantage of this near-far LoD scheme is that it allows for precise control of the rendering load, which is limited to the geometry close to the user. Another advantage is that near-far LoD allows scalability with the number of VR application users. This is because the same edge server can compute, cache, and transmit (using multicast where supported) the same far region representation to users who are nearby in the virtual environment. 

\begin{figure}[htb]
        \includegraphics[width=\columnwidth]{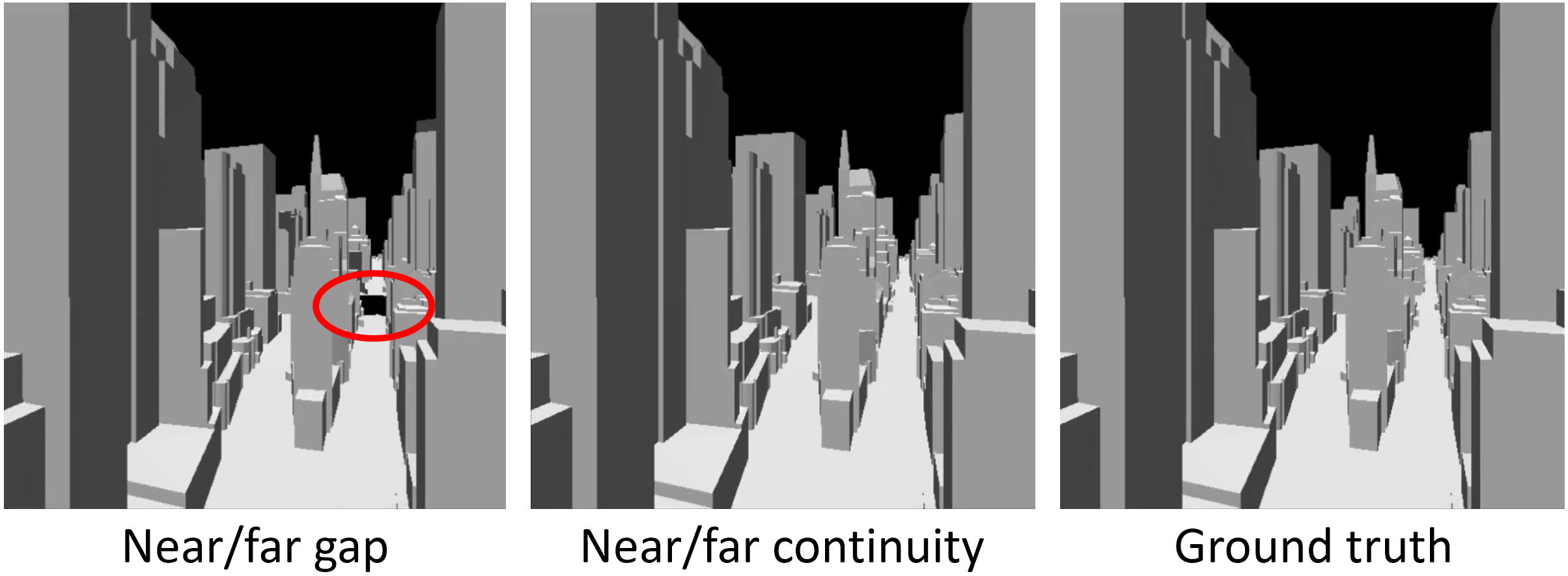}
        \caption{\textcolor{black} {The discontinuity between the near and far regions (left) is eliminated with an intermediate region morphed to connect the two regions (middle), yielding frames comparable to ground truth frames rendered from the geometry of the entire virtual environment (right)~\cite{popescu2022ismar}.}}
        \label{fig:nearFar3way}
\end{figure}

Challenges of near-far LoD schemes include lack of visual continuity between the near and far regions, lack of support for dynamic far regions, and lack of support for motion parallax in the far region, i.e., far region objects at different depths do not move with different speeds in the frame as the user FoV moves. These challenges are being addressed by current research~\cite{popescu2022ismar}, as seen in Figure~\ref{fig:nearFar3way}.

\begin{figure*}[htb]
\centering
        \captionsetup{position=below,skip=4pt}
        \begin{minipage}[htb]{.425\linewidth}
            \includegraphics[width=\columnwidth]{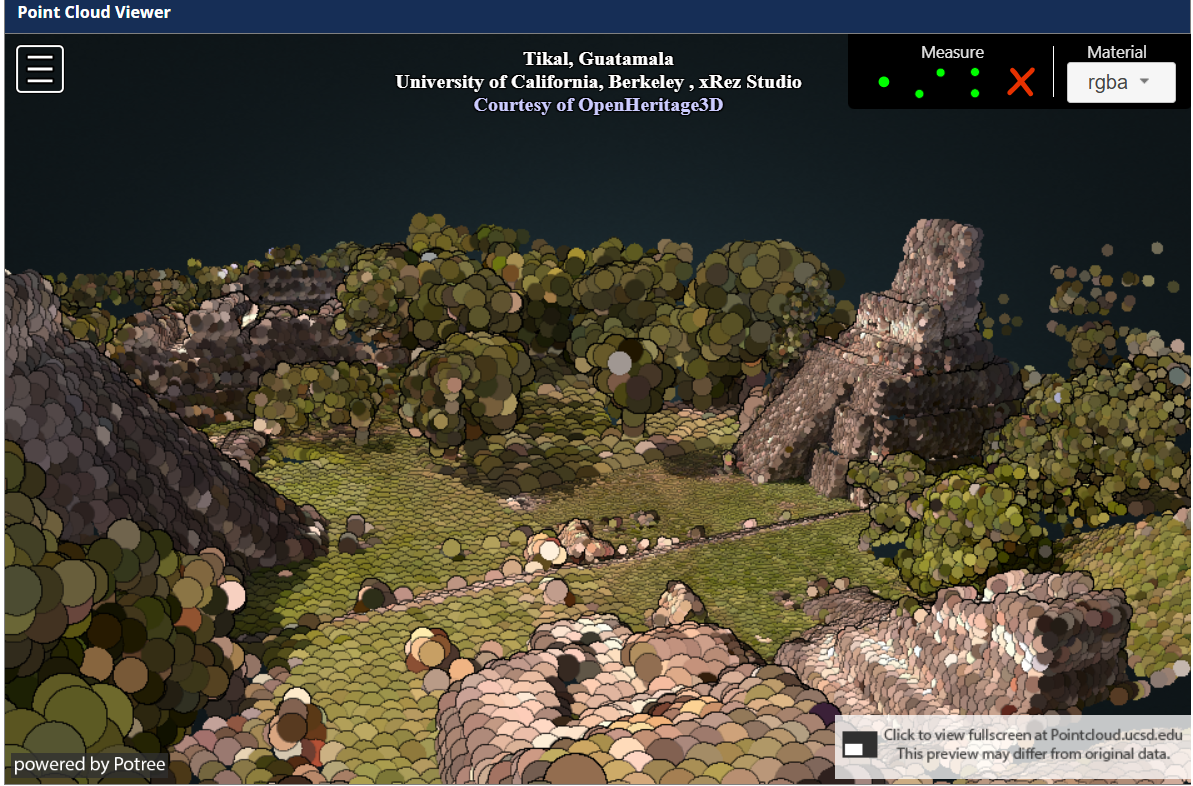}
        \end{minipage}
        \begin{minipage}[htb]{.425\linewidth}
            \includegraphics[width=\columnwidth]{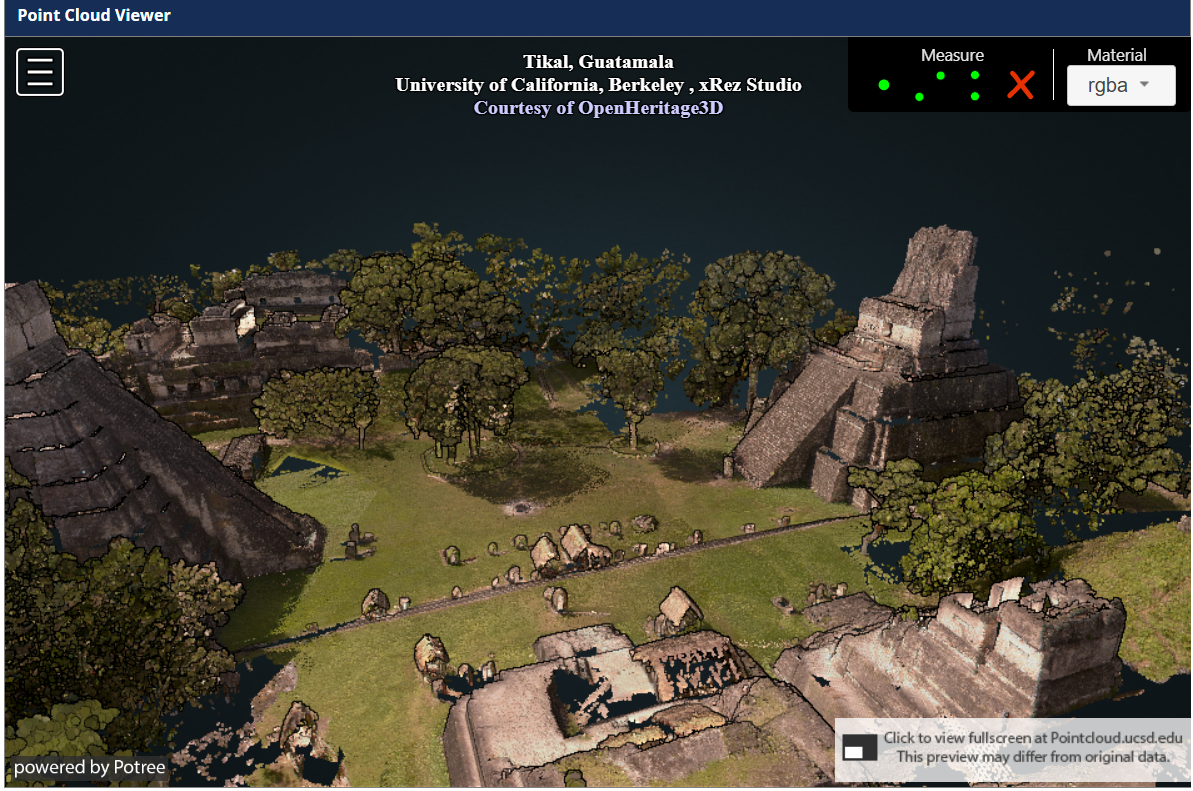}
        \end{minipage}
        \caption{Progressive refinement of point cloud rendering of ``Tikal'' with an increased level of detail in the right image (screen shots obtained from https://doi.org/10.26301/708h-ss96 from openheritage3d.org).}
        \label{fig:ptcloud}
\end{figure*}

\textbf{Example 2:} Consider LoD adaptation of point clouds. The uniformity of the explicit representation provided by point clouds, static or dynamic, considerably simplifies LoD adaptation~\cite{vivo,GROOT20}. Storing the points in 3D hierarchical space partitioning schemes, such as octrees \textcolor{black}{(where each internal node has exactly eight children)} or kd-trees \textcolor{black}{(binary trees where every node is a k-dimensional point)}, provides the LoDs needed for specific viewing distances, network conditions, cache configurations, or client rendering capabilities.
For example, Figure~\ref{fig:ptcloud} illustrates progressive refinement when rendering a point cloud with a coarse LoD (left) to a fine LoD (right). 

\textbf{Example 3:} Consider visibility-based adaptation of 360$^\circ$ panoramic videos~\cite{flare,rubiks,pano}. In this case, the visibility problem is reduced to view frustum culling. The key challenges in this case are accurate user view prediction, robustness to view prediction error, and cropping the 360$^\circ$ frame to its intersection with the view frustum.
View prediction takes a number of approaches, from extending the FoV of the current frame in all directions or in the direction of the current view rotation, to leveraging application knowledge regarding view directions of interest, to mining or learning from previous user traces. Robustness to view prediction error is achieved by padding the predicted view. 

Isolating the part of the frame inside the view frustum can be done by splitting the 360$^\circ$ into rectangular tiles and selecting the tiles that overlap with the predicted view frustum. To avoid the overhead of managing a large number of tiles, large tile sizes are used, e.g., the 360$^\circ$ frame is split into 6 $\times$ 4 tiles~\cite{flare}. As modern headsets have large FoVs of over 90$^\circ \times$ 90$^\circ$, and since an equirectangular frame has a 360$^\circ \times$ 180$^\circ$ FoV, theoretical savings from view frustum culling have an upper limit of 8. Factoring in the need to cover several seconds of user view direction rotation, the padding for view prediction error robustness, and the coarseness of the tiles, the bandwidth savings are eroded. An alternative to tiling is using a non-linear sampling of the 360$^\circ$ frame, encoding the predicted FoV of view at higher resolution, and padding it with a decreasing resolution for view prediction error mitigation~\cite{CoREpaper}. Non-linear sampling reduces bandwidth requirements at the cost of storage at the server.

\section{Prefetching and Caching}

Stored video streaming masks variability in inter-frame timing through prefetching and playout buffering. The variability in timing is caused by dynamic network conditions that may occur due to wireless interference and the user's physical mobility. Once video content is displayed on the client device, it can be discarded from the client buffer. 

In contrast to video streaming, data describing the static parts of a virtual environment can be downloaded ahead of time and be reused, depending on how the user navigates through the virtual environment. When a client has limited storage, energy, or communication capacity, it must decide what data to prefetch and what to keep in its cache to achieve the most efficient usage of available resources while maintaining acceptable QoE. Once a user teleports and starts navigating (and potentially backtracking) through a virtual environment, data requirements of that user will be highly correlated and significantly benefit from caching (even if FoV is continually shifting).

If data representing a virtual environment is only available in the cloud, it may be useful to selectively bring it closer to the clients leveraging edge compute and caching resources. As discussed earlier, doing so allows possible (partial) offloading of rendering workload as well as more rapid delivery of urgently needed data to clients. 
Content delivery to clients can be done proactively (predictively) or reactively to demand.
Further, if clients are both physically nearby, e.g., share the same edge resources, and nearby in the virtual environment, they can benefit from shared caching at the edge, to store and deliver data of common interest. In this case, multicast transmission can deliver data that clients currently require or may soon require~\cite{M5,MuV2}. 

Shared edge resources are costly and relatively scarce. Thus, questions arise as to their fair use among clients. The problem can be defined as fair caching of prefetched content based on forecasting mobility in the virtual environment. A per-user (or user-group) attribution of such content in the edge cache can inform fair caching decisions, where attribution can be ``discounted'' if the content is distributed to other users over its lifetime in the cache or if the content resides in ``colder'' portions of the cache. Ideas of multi-resource fairness~\cite{DRF} can be applied to reckon fair use of different types of resources including across heterogeneous edge servers. In this context, a GPU can dynamically scale-down the LoD of an object to reduce the load on limited resources. As edge resources are stressed, the edge can naturally focus on activities that have maximal communal benefit, e.g., rendering key objects, particularly at lower LoDs, and user-to-virtual environment or user-to-user interactions.

Caching in VR therefore involves finding a good middle ground between proactively storing data that may be of interest in the future and keeping data capturing the virtual environment already explored in case it is revisited. Further, as mentioned earlier, the availability of multiple LoDs for parts of a virtual environment enables quickly delivering a first cut of required data that can be progressively refined with an associated progressively improved quality~\cite{Bovik11}. The caching policy will be involved in making tradeoffs between quality and between content that may be reused by a particular client or a group of clients.

\section{Transport}

Video streaming applications use the DASH protocol operating on top of standard HTTP servers. HTTP runs on top of TCP or on top of the QUIC reliable stream service~\cite{quic2017}. In both cases, video is delivered \emph{reliably} and \emph{in order} to the application. Since video is encoded using standards such as MPEG before being packetized (based on the network maximum transmission unit), it must be delivered both reliably and in order so that the application can decode it and display it.

VR data types are more heterogeneous than video, and hence require different transport methods based on the characteristics of the particular data type. For example, a user's head pose (position and orientation) is encoded with six floats or 24 bytes per frame, it is updated over 70 times per second, and an obsolete head pose is not useful, so its re-transmission should be forgone in favor of sending the current pose. \textcolor{black}{UDP or the unreliable datagram extension to QUIC may be more suitable than TCP or QUIC reliable streams for transmitting a head pose.}

In contrast to a user's head pose, a large building close to the user in an urban virtual environment covers a large part of the user's FoV and has a long useful life; hence, it should be transferred reliably. However, TCP or QUIC reliable streams are also not ideal for transmitting this data. In-order delivery in TCP and in QUIC streams creates a \emph{head-of-line blocking} problem in case of network packet loss. This is because data received after a loss is not delivered to the client VR application --- even if it can be independently decoded and rendered --- until the lost data is retransmitted by the server and received by the client. Unreliable UDP does not suffer from this head-of-line blocking problem, but leaves it to the application to request retransmissions of lost packets. 

VR applications could benefit from a transport service that requests retransmissions, but delivers out-of-order data to the application. This service would be ideal for transmitting objects such as the large building close to the user discussed above.
The application can immediately decode and render any arriving data, e.g., points in a point cloud or triangles in a triangle mesh, and need not wait for retransmitted data to arrive.

The nearby virtual environment objects need reliable transmission (at least at a low LoD) since they have a long useful life and may be needed any time during navigation before the user teleports. It may, however, be useful to allow an application to cancel already-requested transmissions if they have become obsolete, e.g., because a user has teleported to a different location in the virtual environment and no longer needs this object. This cancellation feature is supported by modern protocols such as QUIC~\cite{quic2017}.

Prioritizing the scheduling of transmissions of virtual environment data that is currently visible or will soon be visible is critical so that the client can render it as soon as it arrives. Prioritizing low LoD over high LoD is also often advantageous when progressive refinement is used. In multi-user settings where a set of users is being served by the same edge server, multicast data~\cite{M5,MuV2} that benefits multiple users is clearly of higher scheduling priority than data benefiting a single user. 

\section{Open Research Problems}

\textcolor{black}{We identify three high-priority research thrusts that are needed to support the delivery of high-quality VR content to users who are physically mobile, e.g., users walking through cityscapes or in moving vehicles. Research along these thrusts is important to enable the design of rendering, adaptation, prefetching, caching and transport strategies, especially when leveraging the support of an edge server or cloud.}

\textcolor{black}{\textbf{QoE measure.}
One or more suitable QoE measures need to be defined for VR content. For example, over a period of time, a QoE measure could be based on comparing the subset of rendered objects and associated levels of detail in the users' FoVs to an idealized scenario where all visible objects are correctly rendered and instantaneously appear in the FoVs upon demand.
Such comparisons can be averaged over time and across different users and can be weighted to capture the visual importance of objects that are nearby versus further away.
In particular, such a QoE measure needs to capture the potential for cybersickness.
Moreover, depending on the application, some objects may be particularly important and there may be additional inter-user synchronization issues, e.g., certain objects should be rendered in the FoVs of certain users. The development of QoE measures that accurately capture the tradeoffs that need to be made in delivering VR content represents a major barrier to making systematic progress.} 

\textcolor{black}{\textbf{Levels of detail.}
The availability of different LoDs of objects in a virtual environment enables navigation of key tradeoffs among user QoEs and shared resources (e.g., at an edge server used by caching, rendering and networking functions) for a population of users who are both physically nearby (e.g., in the same 5G cell) and are using the same VR application.  Through the use of different LoDs, a more graceful degradation of QoE may be possible when delays between an edge server and mobile users significantly and unexpectedly increase, which will be critical to avoiding cybersickness.}

\textcolor{black}
{\textbf{Benchmarking.}
A crucial step to advancing the design of VR systems is the ability to realize reproducible performance evaluation based on a set of representative VR benchmark test scenarios.
By definition, the test scenarios need to be adopted by the broader research community to be deemed ``benchmarks.'' For example, the FoV motion of multiple users in a specified VR environment can be simulated with up to 6 DoF per user, including teleportation, correlated motion among sets of users, and both the ``instantaneous'' FoV and a predicted FoV (where the method of prediction is also researcher-specified). Realistic benchmarks can be obtained by involving human subjects using real VR applications in a variety of system and network settings.}

\section{Conclusions}
Although virtual reality has the potential for broad societal impact, realizing this potential will require innovations in system and networking support. Solutions that have been developed for video applications are not directly applicable to virtual reality. This is because virtual reality is immersive, with highly interactive and unpredictable multi-user real-time experiences taking place in a heterogeneous virtual environments. In addition to developing accurate quality of experience metrics \textcolor{black}{and common benchmarks}, major innovations are required in the rendering, adaptation, prefetching, caching and transport mechanisms in order to support future virtual reality applications.

\section*{Acknowledgments}
This work has been supported in part by collaborative NSF grants 2212200, 2212201, and 2212202. 

\bibliographystyle{plain}
\bibliography{main,gk}
\end{document}